\documentclass[seceq]{ptptex}

\usepackage{graphicx}
\usepackage{wrapft}

\title{
Computational aspects of nuclear coupled-cluster theory
}

\author{
Gaute \textsc{Hagen}$^{1,2}$
and Hai Ah \textsc{Nam}$^{3}$
}

\inst{
$^1$Department of Physics and Astronomy, University of Tennessee
Knoxville, TN 37996, USA \\
$^2$Physics Division, Oak Ridge National Laboratory, P.O. Box 2008,
Oak Ridge, TN 37831, USA \\
$^3$ National Center for Computational Sciences Division, Oak Ridge Leadership Computing Facility, Oak Ridge National Laboratory, P.O. Box 2008, Oak Ridge, TN 37831, USA
}

\recdate{
March 16, 2012}

\abst{
We discuss computational aspects of the spherical coupled-cluster method
specific to the nuclear many-body problem. Using chiral
nucleon-nucleon interaction at next-to-next-to-next-to leading order
(N$^3$LO) with cutoff $\Lambda = 500$MeV, we present coupled-cluster
results for the ground state of $^{40}$Ca. Scaling and performance studies are
presented together with challenges we meet with when extending
the coupled-cluster effort to nuclei mass hundred and beyond.}

\begin{document}
\maketitle

\section{Introduction}
The low-energy nuclear many-body problem is a challenging undertaking,
however, in the last decade there has been significant progress in first
principle calculations of nuclei \cite{Piep01,navratil09, DeanLee, koshiroh11,
roth09, barbieri09, fujii09, hagen10, maris09}.
With the advance of chiral effective field theory
\cite{Weinberg,Kolck94,Epel00,N3LO,N3LOEGM,machleidt02} which allows for a systematic and
consistent derivation of the nuclear forces rooted in low-energy
Quantum-Chromo-Dynamics (QCD), and with the development of advanced
many-body methods and high performance computing facilities,
first principle computations of medium mass and neutron rich nuclei
at the extremes of the nuclear chart are now within reach.
The computational cost involved in these calculations grows rapidly as
one moves beyond the lightest towards the medium mass region and
beyond, and it has been crucial to implement techniques that scale
gently with system size and code development that leverages the benefits of modern
architecture at high-performance computing facilities.

The coupled-cluster (CC) method is an optimal approach to medium mass and
neutron rich nuclei as it is an ideal compromise between accuracy on
the one hand and computational cost on the other.  Coupled-cluster
theory was introduced in nuclear physics in the late 1950's by Coester and
K{\" u}mmel \cite{Coe58, Coe60} and was shortly thereafter introduced in
quantum chemistry by {{\v C}{\'\i}{\v z}ek} \cite{Ciz66,Ciz69}.
Coupled-cluster theory has now been established as the ``gold standard'' for
first principle computations in quantum chemistry, see Ref.~\cite{Bar07}
for a recent review. Only in the last decade has
coupled-cluster reemerged in the nuclear physics community and
has established itself as a state-of-the art approach to structure
of medium mass and neutron rich nuclei
\cite{Hei99, Mi00,Dean04,Kow04,Hag08,jansen11,hagen10b, roth12}.

The nuclear coupled-cluster code suite developed at Oak Ridge
National Laboratory, called NUCCOR (Nuclear Coupled-Cluster - Oak Ridge)
has been further advanced under UNEDF (Universal Nuclear Energy Density Functional)
a five-year SciDAC
(Scientific Discovery through Advanced Computing) project \cite{UNEDF,NPN}.
Both $m$-scheme (NUCCOR) and $j$-coupled (spherical)
(NUCCOR-CCSD(T)) schemes for implementing CC at various levels of
clustering approximation (i.e. single, doubles, triples, etc.) has been
developed. Under UNEDF, considerable effort has been put into optimizing the
NUCCOR-CCSD(T) V1.0 doubles kernel for the Jaguar Cray XT5
system, Oak Ridge Leadership Computing Facility's (OLCF) flagship leadership computing resource \cite{exascale}.

To continue to achieve breakthrough discoveries in low-energy nuclear
physics using CC, we need to push investigations to larger medium-mass
nuclei in larger model spaces.  This requires continued emphasis on
scaling the CC algorithms and performance optimization.  Additionally,
Jaguar is presently undergoing a multi-phase upgrade to become Titan,
a Cray XK6 with a single socket of AMD's 16-core interlagos chip with
32 GB of shared memory and another socket with the NVidia GPUs on a single node.
Therefore we need to further develop and improve our implementation of
coupled-cluster theory in preparation for the architectural change.

This paper is organized as follows.  In section 2 we provide an
overview of the spherical coupled-cluster method used in NUCCOR-CCSD(T) and present
convergence properties for the ground state of $^{40}$Ca.
Section 3 describes the computational considerations of the
CC code suite and presents performance results on Jaguar
XK6 without GPUs, and describes future
code developments necessary for continued research. Section 4 provides
conclusions and outlook.

\section{Nuclear coupled-cluster theory}
In coupled-cluster theory we write the $A$-nucleon ground state wave
function $\Psi_0$ in the following form
\begin{equation}
\vert \Psi_0\rangle =  e^{-\hat{T}}|\phi_0\rangle, \:\:\:\:\:\: \hat{T} = \hat{T}_1 + \hat{T}_2 + \ldots + \hat{T}_A.
\label{Psi}
\end{equation}
Here $|\phi_0\rangle$ is the uncorrelated (mean-field) reference
state, and $\hat{T}$ is a linear expansion in particle-hole
excitations operators.
The $k$-particle $k$-hole ($k$p-$k$h) cluster operator is
\begin{equation}
\hat{T}_k =
\frac{1}{(k!)^2} \sum_{i_1,\ldots,i_k; a_1,\ldots,a_k} t_{i_1\ldots
i_k}^{a_1\ldots a_k}
\hat{a}^\dagger_{a_1}\ldots\hat{a}^\dagger_{a_k}
\hat{a}_{i_k}\ldots\hat{a}_{i_1} \,.
\label{T}
\end{equation}
Here and in the following, $i, j, k,\ldots$ label occupied single-particle
orbitals while $a,b,c,\ldots$ label unoccupied orbitals.
From the exponential wave-function ansatz in Eq.~(\ref{Psi}) it follows
directly that coupled-cluster theory is based on the similarity transform
\begin{equation}
\label{hbar}
\overline{H_N}=e^{-\hat{T}} \hat{H_N} e^{\hat{T}}
\end{equation}
of the normal-ordered Hamiltonian $\hat{H_N}$. Here, the Hamiltonian is
normal-ordered with respect to a reference state $|\phi_0\rangle$.
The most commonly used approximation is coupled-cluster with
singles-and-doubles excitations (CCSD) where $\hat{T}\approx \hat{T}_1+\hat{T}_2$.
The unknown amplitudes $t_i^a$ and $t_{ij}^{ab}$
in Eq.~(\ref{T}) are determined from the solution of the coupled-cluster equations
\begin{eqnarray}
\label{ccsd}
0 &=& \langle \phi_i^a | \overline{H} | \phi_0\rangle \,, \\
0 &=& \langle \phi_{ij}^{ab} | \overline{H} | \phi_0\rangle \,.
\end{eqnarray}
Here $|\phi_i^a\rangle = \hat{a}_{a}^\dagger\hat{a}_{i}|\phi_0\rangle$
is a 1p-1h excitation of the reference state, and
$|\phi_{ij}^{ab}\rangle$ is a similarly defined 2p-2h excited
state. The CCSD equations~(\ref{ccsd}) thus demand that the reference
state $|\phi_0\rangle$ has no 1p-1h and no 2p-2h excitations, i.e. it is
an eigenstate of the similarity transformed Hamiltonian~(\ref{hbar})
in the space of all 1p-1h and 2p-2h excited states. Once the CCSD
equations are solved, the ground-state energy is computed as
\begin{equation}
E = \langle \phi_0 | \overline{H} | \phi_0\rangle \,.
\end{equation}
Using Wick's theorem the coupled-cluster equations in~(\ref{ccsd}) can
be written as a set of coupled non-linear set of equations in the
$t_i^a$ and $t_{ij}^{ab}$ amplitudes. However, this task is
very tedious and error prone, and a much more direct and intuitive
derivation procedes via a diagrammatic derivation using a set of well
defined diagram rules (see e.g. Ref.~\cite{Bar07}).
In the CCSD approximation the number of non-linear equations are
given by $n_on_u + n_o^2n_u^2$. Here $n_o$ is the number of occupied
orbitals in the reference state and $n_u$ is the number of unoccupied
states above the fermi level. In a typical case calculating the
CCSD ground state of $^{40}$Ca in 15 major oscillator shells we have
a total of 40 occupied and 2680 unoccupied orbtials, resulting in a
total of $\sim 10^{10}$ non-linear coupled equations. This
problem is too large to be solved by direct inversion techniques,
and we use an iterative solution scheme together with krylov subspace methods
such as Broyden \cite{johnson88, mario08} or Direct-Inversion in the Iterative
Sub-Space (DIIS) \cite{pulay80} as convergence accelators.

The most expensive contribution to the $T_2$ amplitude in the
CCSD approximation is
\begin{equation}
t_{ij}^{ab} \leftarrow \sum_{cd}\langle ab | \chi | cd \rangle t_{ij}^{cd},
\label{eq:t2eqn}
\end{equation}
it is clear that the computational cost associated with this
contribution is $n_o^2n_u^4$.
Coupled-cluster theory therefore scales polynomial with the system
size, which is a rather soft scaling compared to the combinatorial
scaling of methods such as the full configuration interaction (FCI)
method. Equation~(\ref{eq:t2eqn}) can be rewritten in a matrix-matrix
multiplication form, so
that we can utilize the optimized basic linear algebra library
(BLAS) \cite{blas}.
Although the scaling of the CC method is rather
soft, it is clear that due to memory limitations we quickly reach the
limit of the maximum model space that can be handled on modern
computers. For example, in 15 major oscillator shells the two-body
interaction alone would require about 600 TByte of memory alone and
the $T_2$ amplitudes would require 100 GByte of memory.

\begin{figure}[h]
\begin{center}
\includegraphics[bb=0 0 650 650,clip,trim=0cm 0cm 0cm 0cm,width=0.35\textwidth]{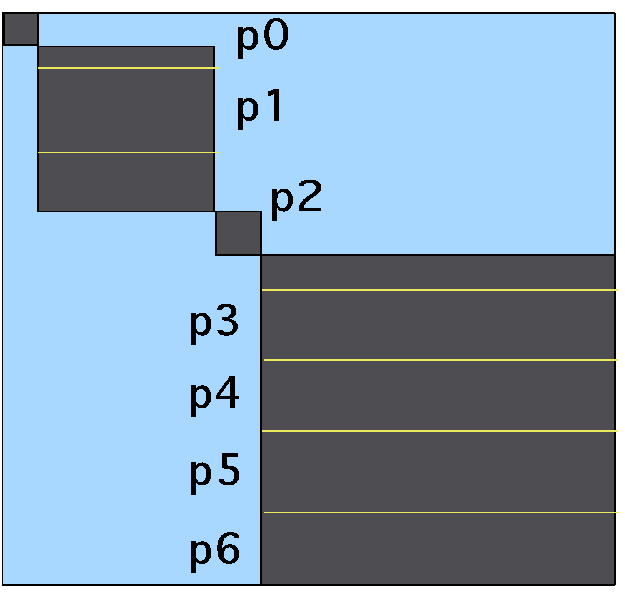}
\caption{(Color online) Block diagonal structure of the interaction
  and parallel distribution scheme}
\label{fig3}
\end{center}
\end{figure}

In order to overcome this bottleneck, we recently derived and
implemented the coupled-cluster equations in a
coupled-angular momentum scheme \cite{Hag08, hagen10}, thereby
utilizing the spherical symmetry of closed shell nuclei.
For such nuclei, the cluster operator in Eq.~(\ref{T}) is a scalar under rotation, and depends
only on reduced amplitudes.
The reduced matrix elements of the Hamiltonian are stored in six different
blocks according to
\[
\label{hdist}
\left[ pppp \right], \left[ hppp \right], \left[ hhpp \right],
\left[ hphp \right], \left[ hhhp \right], \left[ hhhh \right],
\]
where $p$ denotes particle states and $h$ denotes hole states.
These matrices are sparse and block diagonal, each block is
defined by a set of quantum numbers $J^\pi, T_z$. The largest matrices
are the $\left[ pppp \right]$ and  $\left[ hppp \right] $ are
distributed among the processors by adding row by row to given
processor until a criterion for optimal load balancing and
memory distribution is reached, as illustrated in Fig.~\ref{fig3}
By this distribution scheme each processor will have a set of square
and rectangular matrices for a subset of channels (quantum numbers)
 $J^\pi, T_z$. By this scheme, we can utilize the optimized linear
algebra libraries Blas and Lapack \cite{blas} for matrix-matrix and matrix-vector
operations. It is clear that the similarity transformed Hamiltonian
is also a scalar under rotation and can be distributed in the same
scheme, and it is straight forward (but somewhat tedious) to work out
the CCSD equations within this formulation.

As an example of the computational savings we achieve by switching
from an uncoupled to a $j$-coupled scheme, we consider the case
of $^{40}$Ca in 15 major shells. In $m-$scheme we have a total of
2720 orbitals while in the coupled $j$-scheme we have 240 orbitals,
the number of non-linear CCSD equations in $m$-scheme is $\sim 10^{10}$ while
in $j-$scheme we have $\sim 10^6$. The various interaction
blocks in Fig.~\ref{hdist} appear in various topologically different
diagrams in the CCSD equations. Therefore it is very difficult to make the CCSD code
scale optimally with increasing number of processors. We choose to
distribute the blocks according to the number of computational
cycles of the most expensive diagrams they appear in. Therefore
scaling can only be optimal for these subsets of diagrams, while
the remaining ones will scale less optimally.

\begin{figure}[h]
  \includegraphics[width=0.5\textwidth, clip]{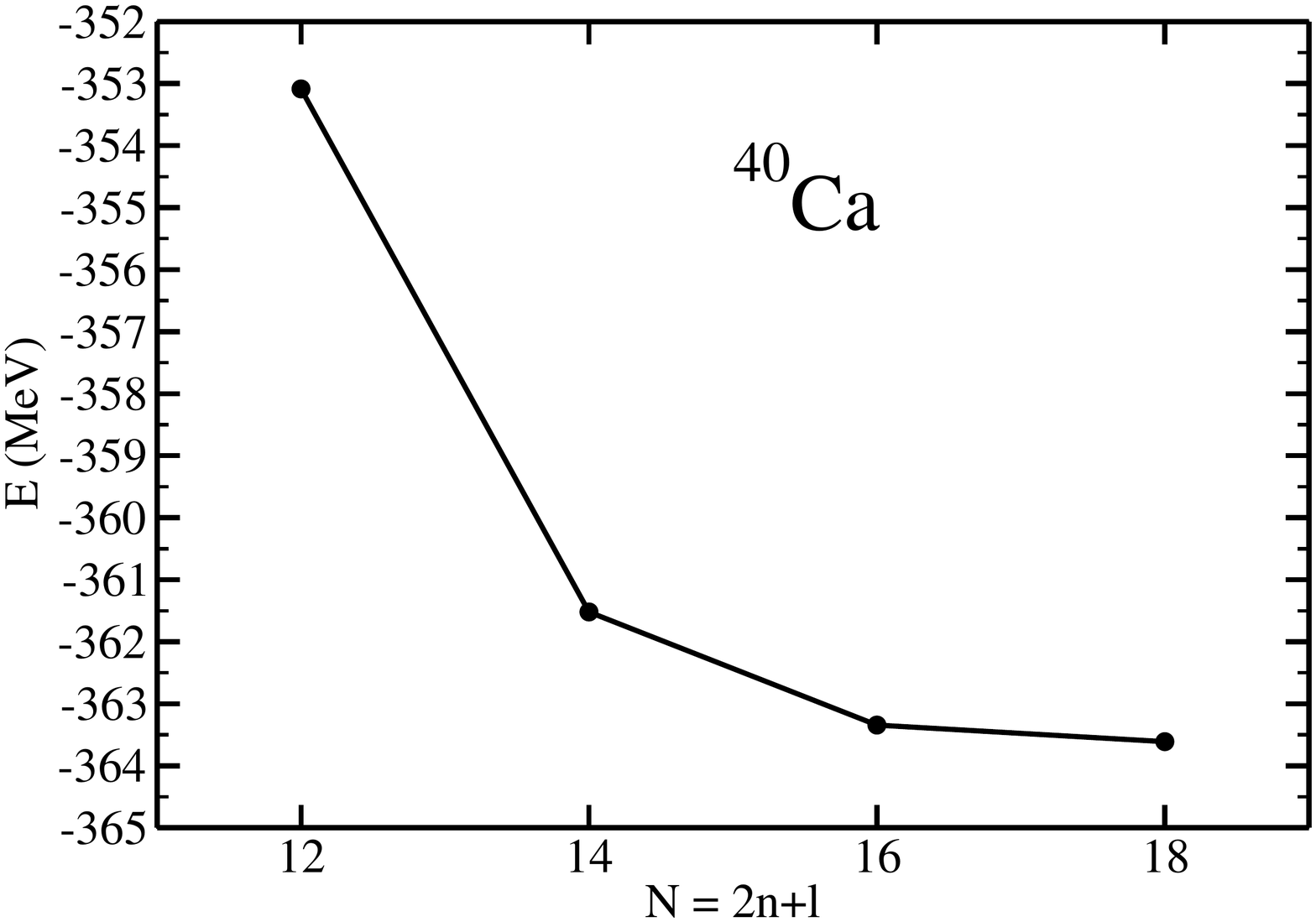}
  \includegraphics[width=0.5\textwidth, clip]{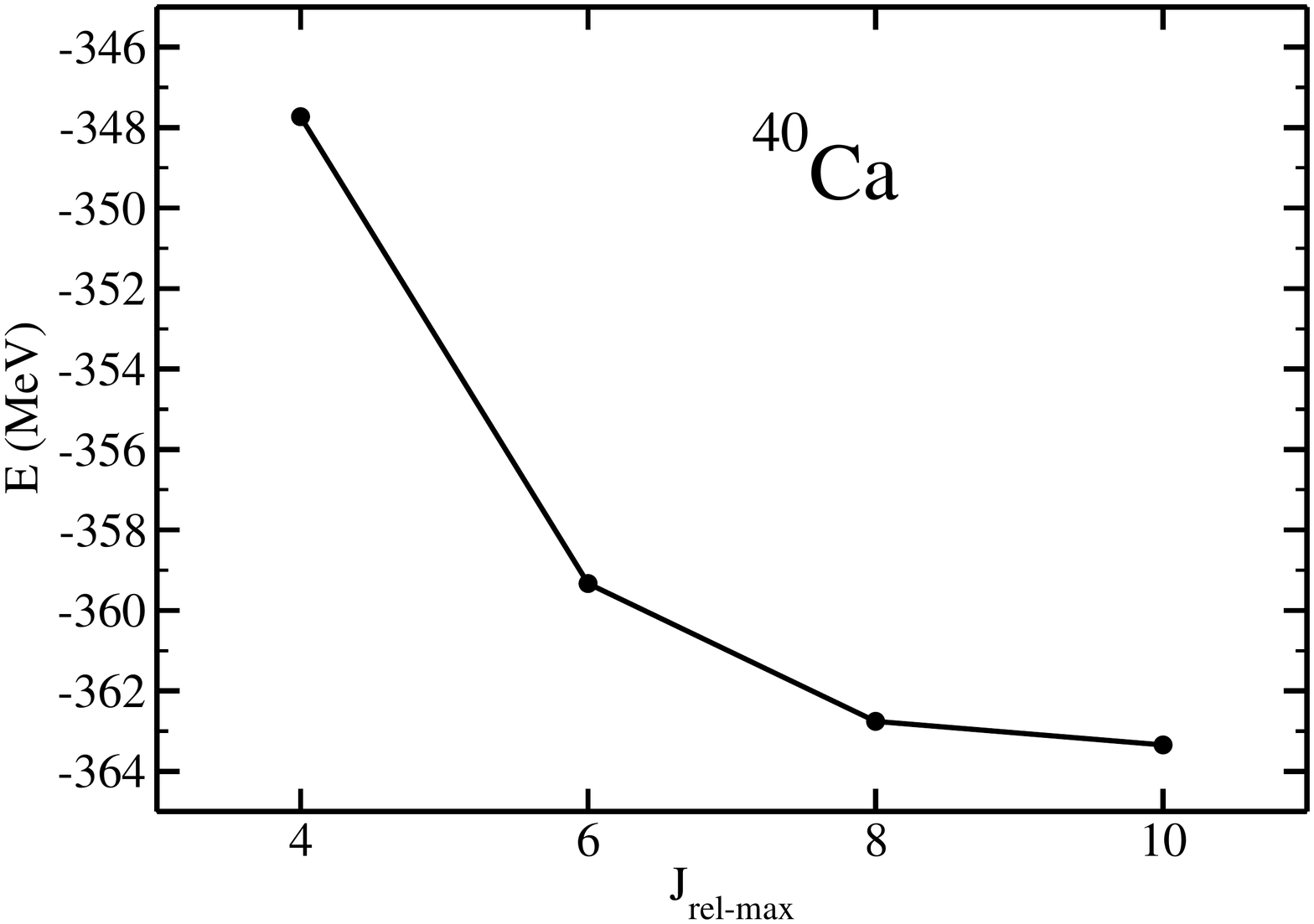}
  \caption{Convergence of $\Lambda$-CCSD(T) ground state energy of
    $^{40}$Ca as a function $\mathrm{N=2n+l}$ (left figure) and
    $\rm{J_{rel-max}}$ (right figure)}
  \label{Ca40_conv1}
\end{figure}

To illustrate the convergence properties of the coupled-cluster
ground state as a function of model space size we compute the ground
state of $^{40}$Ca within $\Lambda$-CCSD(T) approach \cite{hagen10}. We start from the
intrinsic Hamiltonian where the nucleon-nucleon interaction is given by
the N$^3$LO interaction ($\Lambda_\chi = 500$~MeV$c^{-1}$)
by Entem and Machleidt \cite{machleidt02,N3LO}. The nucleon-nucleon interaction
is given in relative coordinates, and in order to express it in a harmonic oscillator
single-particle basis we need to transform it to the laboratory frame using
the Brody-Moshinsky transformation \cite{brody60}.
This transformation depends on the number of partial waves we include
in the relative coordinate frame given, and is given by $J_{\rm rel-max}$.
In order to have fully converged calculations
we need to make sure we have included sufficient number of partial waves
in the Brody-Moshinsky transformation. The convergence is therefore two-fold,
(i) as a function of the single-particle basis size, and (ii) as a function of number
of relative partial waves included in the relative-to-laboratory frame transformation.
We perform our calculations using a Hartree-Fock basis expressed in a harmonic oscillator
single-particle basis with a frequency of $\hbar\omega=26$~MeV.
In Fig.~\ref{Ca40_conv1} we show the convergence for the ground state
of $^{40}$Ca as a function of number of harmonic oscillator shells $N = 2n+l$
(left figure), and as a function of $J_{\rm rel-max}$ for the Brody-Moshinsky transformation,
keeping the single-particle space fixed at $N=16$ shells (right figure).
We see that $N=16$ is sufficient size for the single-particle model space
in order to converge the ground state to within 1~MeV, while
we need to include relative partial waves up to $J_{\rm rel-max}=8,10$ to reach
convergence for the Brody-Moshinsky transformation. In
our earlier calculations \cite{Hag08, hagen10} we used
$J_{\rm rel-max}= 4$, and therefore underestimated the binding
energy of $^{40}$Ca by $\sim 15$~MeV.

\section{Computational Considerations}

In coupled-cluster calculations, the computational challenge consists of scaling the
computational resources with the increasing size of the model
space and size of nuclei. For an interaction with momentum cutoff $\Lambda$ and a nucleus
with radius $R$, the size of the single-particle basis scales as
$(R\Lambda)^3\sim A\Lambda^3$ \cite{hagen10}, with $A$ being the mass number. In 20
oscillator shells, the matrix elements of the two body-interaction
require about 100~GByte of memory, and for tin isotopes, we expect
that 25-30 shells will be required for present-day chiral
interactions.

The NUCCOR-CCSD(T) application has been developed for the Jaguar supercomputer located at Oak Ridge National Laboratory.  It utilizes MPI and threaded BLAS and LAPACK libraries, suited for the Cray XT5 architecture comprised of dual hex-core processors with 16 GB of memory on the node.  Jaguar is presently undergoing a phased upgrade from an XT5 to an eventual Cray XK6, called Titan, with a single interlagos 16-core chip with 32 GB of memory paired with NVidia GPUs.  The current phase of the upgrade has fully deployed the CPU enhancements with the GPU additions expected to complete at the end of 2012.

Hybrid programming, using both MPI and OpenMP, is more advantageous in the current configuration of the system due to the large shared memory and the large number of cores on a single node.  In an effort to efficiently utilize this new architecture, we have made added further parallelism in the triples calculation using OpenMP.  This section presents performance results of the improved NUCCOR-CCSD(T) V2.0 application on the current Jaguar XK6 configuration, without GPUs.

\subsection{Scaling by model space}
\begin{figure}[h]
\begin{center}
\includegraphics[bb=0 0 750 700,clip,trim=0cm 0cm 0cm 0cm,width=0.60\textwidth]{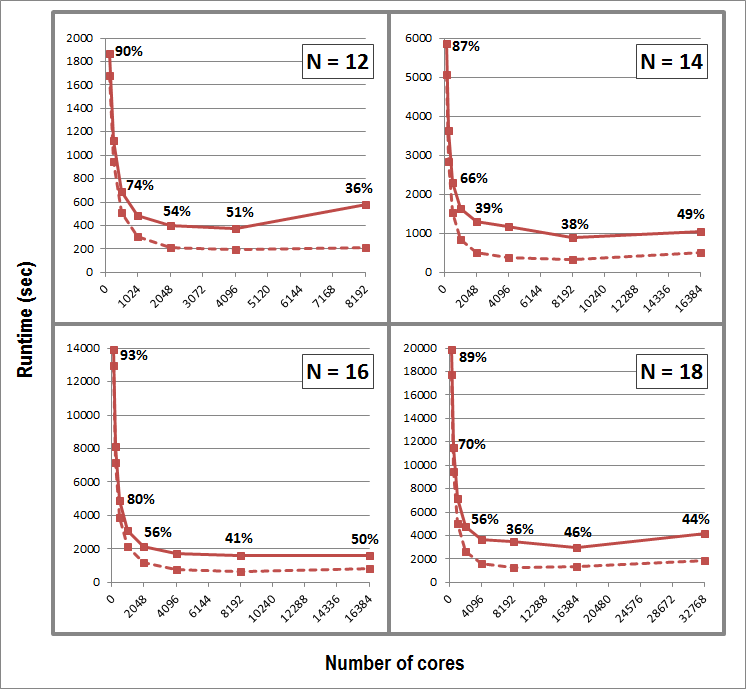}
\caption{(Color online) Strong scaling results for $^{40}$Ca in model spaces $N = 12$, $N = 14$, $N = 16$, and $N = 18$.  Dashed lines show triples runtime only and solid lines represent total runtime.  Percentage labels show percent of total runtime spent in triples calculation.}
\label{scaling}
\end{center}
\end{figure}

Figure~\ref{scaling} shows strong scaling results for NUCCOR-CCSD(T) V2.0 using the PGI 11.10.0 compiler with optimization \emph{-fast} and the Cray LibSci 11.0.4 scientific threaded library for $^{40}$Ca in model spaces $N = 12, 14, 16$ and $18$.  Runtimes for the triples calculation are presented with total runtimes along with labels showing the percentage of runtime used in the triples calculation at various numbers of cores.  These calculations utilize 4 MPI processes each spawning 4 threads using OpenMP and the threaded libraries to optimally use all cores on a node.

Seen in all model spaces, the runtime of NUCCOR-CCSD(T) V2.0 is dominated by the triples calculation at small numbers of processors, consuming roughly $90\%$ of the total runtime.  Utilizing more processors shows a dramatic decrease in the runtime of the triples calculation, which in turn decreases the overall runtime.  At larger numbers of processors, the triples calculation consumes less than $50\%$ of the total runtime, indicating a necessity to revisit performance optimizations for other regions of the code, including the single and doubles calculation, setup, and I/O, which consume the remainder of the runtime.

\begin{figure}[h]
\begin{center}
\includegraphics[bb=0 0 825 400,clip,trim=0cm 0cm 0cm 0cm,width=0.65\textwidth]{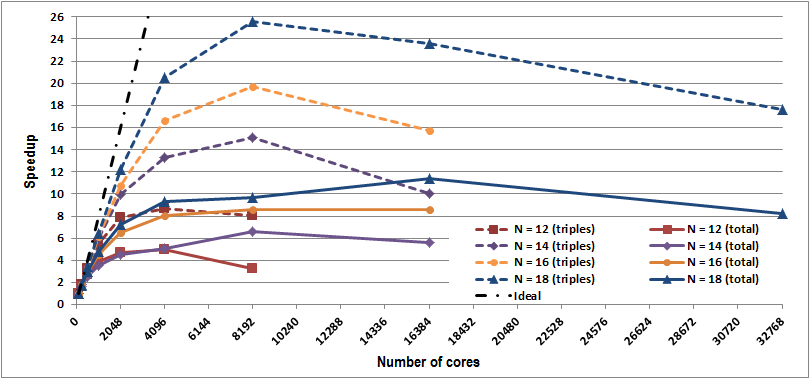}
\caption{(Color online) Speedup from 8 nodes (128 cores) for $^{40}$Ca in model spaces $N = 12$ (red square), $N = 14$ (purple diamond), $N = 16$ (orange circle), and $N = 18$ (blue triangle).  Dashed lines show triples runtime only and solid lines represent total runtime.}
\label{speedup}
\end{center}
\end{figure}
The highest number of cores used in each model space shows the limits of strong scaling for both the triples and the total runtime.  This is further evident in the speedup trends shown in Fig.~\ref{speedup}, where the speedup at each model space drops sharply at the highest processor count for each model space.  We use 128 cores for $T(1)$ in the speedup calculations, which is the smallest configuration to run all model spaces.  Figure~\ref{speedup} shows drastic speedup in the triples calculations with larger number of cores, but that the total runtime only improves by roughly half those values.  With improvements in the triples calculation, which originally dominated the NUCCOR-CCSD(T) runtime, other regions of code utilizing MPI only will also require further parallelization to utilize the new architecture.

\subsubsection{MPI vs. Threading}
\begin{figure}[h]
\begin{center}
\includegraphics[bb=0 0 700 350,clip,trim=0cm 0cm 0cm 0cm,width=0.65\textwidth]{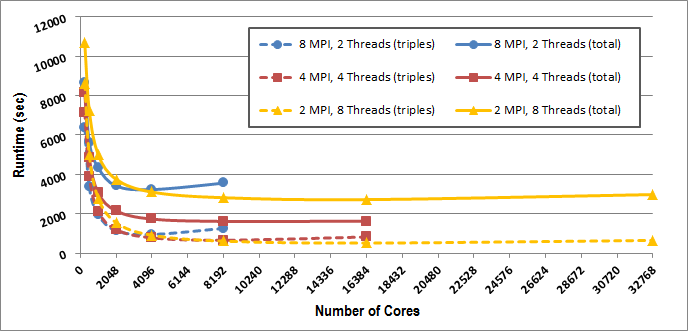}
\caption{(Color online) Comparison of runtimes (in seconds) of NUCCOR-CCSD(T) V2.0 for $^{40}$Ca, $N = 16$ with varying MPI processes and threads.  Dashed lines show triples runtime only and solid lines represent total runtime.  Thread configurations include:  8 MPI processes with 2 threads each (blue dot), 4 MPI processes with 4 threads each (red square), and 2 MPI processes with 8 threads each (yellow triangle) on a node.  All compute cores on the node are utilized.  }
\label{threading}
\end{center}
\end{figure}

Although the use of threads through OpenMP has greatly improved the triples calculation and thus the overall runtime, Fig.~\ref{threading} shows that increasing the number of threads does not improve the total runtime.  In Fig.~\ref{threading} we present scaling results using three configurations, varying the number of MPI processes and threads per process.  We show that using 4 MPI processes and 4 threads performs equally well for the triples calculation as using 2 MPI processes and 8 threads.  Due to the MPI only regions of code, yet to be optimized for the new architecture, the total runtime is worse using less MPI processes and more threads.  Also shown in Fig.~\ref{threading} is a degradation in performance when using more MPI processes and less threads (8 MPI process and 2 threads) since it does not allow the benefits of threading to be realized.

\section{Conclusions and future perspectives}
We have made substantial performance improvements in NUCCOR-CCSD(T) V2.0 by
implementing threading in the triples calculation using OpenMP. For example,
our triples calculation consumed $70-95\%$ of the runtime,
scaling to 8192 cores for $N$=18 in $^{40}$Ca. This example
used 4 MPI processes each spawning 4 threads, resulting in
a total number of 32768 requested MPI processes. Note that we typically
need several runs of this size for each nucleus, in order to determine
convergence and to map out the dependence on the model space parameters.

These improvements now reveal new areas for continued development to
optimize the total runtime performance to reach nuclei larger
than $^{40}$Ca in larger model spaces.  Improvements include further
threading MPI only regions, including the singles and doubles calculations,
setup, and improvements to I/O.

In preparation for utilizing the full capability of Titan, the hybrid CPU-GPU system,
which will be available in Q4 of 2012, we are also going through the exercise of identifying compute-intensive kernels suitable to be off-loaded to the GPU and implementing GPU-optimized libraries, GPU directives, and GPU languages such as
CUDA and OpenCL, as needed to continue to improve the models and implement more realistic system constraints.  For example, the inclusion of continuum effects in weakly bound nuclei and nuclear reactions will require us to utilize a large
number of scattering states which further increases the size of the
model space.  Increasing the order of the coupled-cluster approximation and including higher-body effects, such as three-body and four-body forces, present additional computational challenges to both developers and computing systems.

\section{Acknowledgments}
This work was supported by the Office of Nuclear
Physics, U.S. Department of Energy (Oak Ridge National
Laboratory). This work was supported in part by the
U.S. Department of Energy under Grant No. DE-FC02-07ER41457
(UNEDF SciDAC). This research used resources of the National
Center for Computational Sciences at Oak Ridge National Laboratory,
which is supported by the Office of Science of the U.S. Department of
Energy under Contract No. DE-AC05-00OR22725.


\end{document}